# Localization of DES Supervisory Control with Event Reduction

Vahid Saeidi[1], Ali A. Afzalian[2] and Davood Gharavian[3]
Department of Electrical Eng., Abbaspour School of Engineering,
ShahidBeheshti University, Tehran, Iran
[1] v_saeidi@sbu.ac.ir, [2] Afzalian@sbu.ac.ir, [3] d_gharavian@sbu.ac.ir

**Abstract**

Supervisor localization procedure can be employed to construct local controllers corresponding to component agents in discrete-event systems. The proposed method in [11] is based on state reduction of a monolithic supervisor with respect to each set of controllable events corresponding to each component agent. A supervisor is localizable if state cardinality can be reduced from the reduced supervisor to each local controller. Although event reduction is an important property, the original supervisor localization procedure did not guarantee event reduction in each local controller comparing to the reduced supervisor. In this paper, we propose a method to localize a supervisor with event reduction in each local controller comparing to the reduced supervisor. State reduction facilitates the implementation of local controllers on industrial systems, whereas event reduction reduces communication traffic between each pair of local controllers.

## 1. Introduction

The supervisory control theory (SCT) encounters with some issues such as discrete-event modeling and computational complexity [1]. Modular [2-5] and incremental/compositional [6-9] approaches try to overcome the complexity of the supervisor synthesis. Other approaches tend to reduce the supervisor for simple implementation [10]. Such approaches do not affect on computational complexity reduction.

Supervisor localization procedure, introduced in [11], is a method to distribute the supervisory control of discrete-event systems. This procedure achieves two goals: (i) it preserves the optimality (i.e. minimally restrictive) and non-blocking of the monolithic supervisor, and (ii) it dramatically simplifies each local controller based on the state size criterion. Namely, a supervisor is localizable if the state size of each local controller is less than the state size of the reduced supervisor. Both goals are achieved by a suitable extension of supervisor reduction procedure [10]. This procedure is carried out using control information relevant to the target agent. As the result, each agent obtains its own local controller. However, the control authority of a local controller is strictly local; the observation scope of each local controller is systematically determined in order to guarantee the correct local control action.This procedure cannot guarantee event reduction in each local controller comparing to the reduced supervisor. This procedure has been extended in [12], for localizing the supervisory control under partial observation and using relative observability property [13]. Some applications of supervisor localization can be found in [14].

On the other hand, decomposition of a supervisor [15] is an alternative method to reduce the number of events, considered in decision making by each decentralized supervisor. This method constructs a distributed supervisory control with restricted control authority and restricted observation scope.

In this paper, we propose a supervisor localization procedure which ensures the event reduction from the reduced supervisor to each local controller.

The main common concepts in the supervisor reduction/localization procedures are control consistency of states. A cover on the state set is constructed so that at least a pair of states which are not control consistent with respect to (w.r.t.) all controllable events but are control consistent w.r.t. a subset of controllable events corresponding to a component of the plant belongs to a subset of states which can be lumped. Also, the next states by the same transitions from a pair of states in the plant correspondingly should not be the same.

It is proved that, each local controller which is constructed by the proposed method has less number of events comparing to the reduced supervisor (in the sense of normality [16, 17]). It is an important result. However, event reduction in each local controller does not guarantee decomposability of a supervisor; each local controller needs fewer events for consistent decision making, comparing to the reduced state supervisor. It shows reducing the communication traffic between local controllers in a distributed supervisory control.

An induced generator is constructed by substituting one state for each subset in the control cover. The constructed induced generator must be a deterministic automaton. The set of constructed induced generators is control equivalent to the supervisor w.r.t. the plant.

In Section 2, the necessary preliminaries are reviewed. In Section 3, we propose a method to localize the supervisory control with event reduction. In Section 4, the proposed method is illustrated for supervisory control of industrial transfer line. Finally, concluding remarks and future work are given in Section 5.

## 2. Preliminaries

A discrete-event system is presented by an automaton $\mathbf{G} = (Q, \Sigma, \delta, q_0, Q_m)$, where $Q$ is a finite set of states, with $q_0 \in Q$ as the initial state and $Q_m \subseteq Q$ being the marked states; $\Sigma$ is a finite set of events ($\sigma$) which is partitioned as a set of controllable events $\Sigma_c$, and a set of uncontrollable events $\Sigma_u$, where $\Sigma = \Sigma_c \dot\cup \Sigma_u$. $\delta$ is a transition mapping $\delta: Q \times \Sigma \to Q$, $\delta(q, \sigma) = q'$ gives the next state $q'$ is reached from $q$ by the occurrence of $\sigma$. In this context $\delta(q_0, s)!$ means that $\delta$ is defined for $s$ at $q_0$. $L(\mathbf{G}) \coloneqq \{s \in \Sigma^* | \delta(q_0, s)!\}$ is the closed behavior of $\mathbf{G}$ and $L_m(\mathbf{G}) \coloneqq \{s \in L(\mathbf{G}) | \delta(q_0, s) \in Q_m\}$ is the marked behavior of $\mathbf{G}$ [18, 19]. In this paper, we assume that $\mathbf{G}$ consists of component agents $\mathbf{G}^k$, defined on pairwise disjoint events sets $\Sigma^k$ ($k \in \mathcal{K}$, $\mathcal{K}$ is an index set), i.e. $\Sigma = \dot\cup \{\Sigma^k | k \in \mathcal{K}\}$. Let $L_k \coloneqq L(\mathbf{G}^k)$ and $L_{m,k} \coloneqq L_m(\mathbf{G}^k)$, the closed and marked behaviors of $\mathbf{G}$ are $L(\mathbf{G}) = \| \{L_k | k \in \mathcal{K}\}$ and

$L_m(\mathbf{G}) = \| \{L_{m,k} | k \in \mathcal{K}\}$, respectively, where $\|$ denotes synchronous product [19]. We assume that for every $k \in \mathcal{K}$, $\bar{L}_{m,k} = L_k$ is true. Then $\mathbf{G}$ is necessarily non-blocking (i.e. $\bar{L}_m(\mathbf{G}) = L(\mathbf{G})$). A set of all control patterns is denoted with $\Gamma = \{\gamma \in Pwr(\Sigma) | \gamma \supseteq \Sigma_u\}$. A supervisor of a plant $\mathbf{G}$, is a map $V: L(\mathbf{G}) \to \Gamma$, where $V(s)$ represents the set of enabled events after the occurrence of the string $s \in L(\mathbf{G})$. A pair $(\mathbf{G}, V)$ is written by $V/\mathbf{G}$ and called $\mathbf{G}$ is under supervision by $V$. A behavioral constraint on $\mathbf{G}$ is given by specification language $E \subseteq \Sigma^*$. Let $K \subseteq L_m(\mathbf{G}) \cap E$ be the supremal controllable sublanguage of $E$ w.r.t. $L(\mathbf{G})$ and $\Sigma_u$, i.e. $K = supC(L_m(\mathbf{G}) \cap E)$. If $K \neq \emptyset$, $\mathbf{SUP} = (X, \Sigma, \xi, x_0, X_m)$ is recognizer of $K$. Write $|.|$ for the state size of DES. Then $|\mathbf{SUP}| \leq |\mathbf{G}||\mathbf{E}|$. In applications, engineers want to employ the reduced supervisor $\mathbf{RSUP}$, which has a fewer number of states and is control equivalent to $\mathbf{SUP}$ w.r.t. $\mathbf{G}$ [10], i.e.

$$L_m(\mathbf{G}) \cap L_m(\mathbf{RSUP}) = L_m(\mathbf{SUP}), \quad (1)$$
$$L(\mathbf{G}) \cap L(\mathbf{RSUP}) = L(\mathbf{SUP}). \quad (2)$$

A generator $\mathbf{LOC}^k$ over $\Sigma$, is a local controller for agent $\mathbf{G}^k$, if $\mathbf{LOC}^k$ can disable only events in $\Sigma_c^k$, where $\Sigma_c^k = \Sigma^k \cap \Sigma_c$. Precisely, for all $s \in \Sigma^*$ and $\sigma \in \Sigma$, there holds,

$$s\sigma \in L(\mathbf{G}) \& s \in L(\mathbf{LOC}^k) \& s\sigma \notin L(\mathbf{LOC}^k) \Rightarrow \sigma \in \Sigma_c^k.$$

The observation scope of $\mathbf{LOC}^k$ is not limited to $\Sigma^k$. But, the control authority of a local controller is strictly local [11].

A set of local controllers $\mathbf{LOC} = \{\mathbf{LOC}^k | k \in \mathcal{K}\}$ is constructed, each one for an agent, with $L(\mathbf{LOC}) = \cap \{L(\mathbf{LOC}^k) | k \in \mathcal{K}\}$ and $L_m(\mathbf{LOC}) = \cap \{L_m(\mathbf{LOC}^k) | k \in \mathcal{K}\}$ such that the following relationships hold,

$$L_m(\mathbf{G}) \cap L_m(\mathbf{LOC}) = L_m(\mathbf{SUP}), \quad (3)$$
$$L(\mathbf{G}) \cap L(\mathbf{LOC}) = L(\mathbf{SUP}). \quad (4)$$

We say that, $\mathbf{LOC}$ is control equivalent to $\mathbf{SUP}$ w.r.t. $\mathbf{G}$, if (3) and (4) are satisfied. This formulation is based on state reduction of a monolithic supervisor with respect to disabled controllable events of each component agent.

The natural projection is a mapping $P: \Sigma^* \to \Sigma_0^*$, where (1) $P(\epsilon) := \epsilon$, (2) for $s \in \Sigma^*$, $\sigma \in \Sigma$, $P(s\sigma) := P(s)P(\sigma)$, and (3) $P(\sigma) := \sigma$ if $\sigma \in \Sigma_0$ and $P(\sigma) := \epsilon$ if $\sigma \notin \Sigma_0$. The effect of an arbitrary natural projection $P$ on a string $s \in \Sigma^*$ is to erase the events in $s$, that do not belong to observable events set, $\Sigma_0$. The natural projection $P$ can be extended and denoted by $P: Pwr(\Sigma^*) \to Pwr(\Sigma_0^*)$. For any $L \subseteq \Sigma^*$, $P(L) := \{P(s) | s \in L\}$. The inverse image function of $P$ is denoted by $P^{-1}: Pwr(\Sigma_0^*) \to Pwr(\Sigma^*)$ for any $L \subseteq \Sigma_0^*$, $P^{-1}(L) := \{s \in \Sigma^* | P(s) \in L\}$. $K$ is relative observable w.r.t. $\bar{C}, \mathbf{G}$ and $P$, for $K \subseteq C \subseteq L_m(\mathbf{G})$, where $\bar{K}$ and $\bar{C}$ are prefix closed languages, if for every pair of strings $s, s' \in \Sigma^*$ such that $P(s) = P(s')$, the following two conditions hold [13],

$$(a)(\forall \sigma \in \Sigma)\, s\sigma \in \bar{K}, s' \in \bar{C}, s'\sigma \in L(\mathbf{G}) \Rightarrow s'\sigma \in \bar{K},$$
$$(b)\, s \in K, s' \in \bar{C} \cap L_m(\mathbf{G}) \Rightarrow s' \in K.$$

In the special case, if $C = K$, then the relative observability property is tighten to the observability property. An observation property called normality was defined in [16], that is stronger than the relative observability. $K$ is said to be normal w.r.t. $(L(\mathbf{G}), P)$, if $P^{-1}P(\bar{K}) \cap L(\mathbf{G}) = \bar{K}$, where $L(\mathbf{G})$ is a prefix closed language and $P$ is a natural projection.

## 3. Supervisor localization procedure with event reduction

Let $\mathbf{G} = (Q, \Sigma, \delta, q_0, Q_m)$ be the plant and $\mathbf{SUP} = (X, \Sigma, \xi, x_0, X_m)$ be the monolithic supervisor. Define $E: X \to Pwr(\Sigma)$ as $E(x) = \{\sigma \in \Sigma | \xi(x, \sigma)!\}$. $E(x)$ denotes the set of events enabled at state $x$. Next, define $D: X \to Pwr(\Sigma)$ as $D(x) = \{\sigma \in \Sigma | \neg \xi(x, \sigma)! \,\&\, (\exists s \in \Sigma^*)[\xi(x_0, s) = x \,\&\, \delta(q_0, s\sigma)!]\}$. $D(x)$ is the set of events, which are disabled at state $x$. Define $M: X \to \{1,0\}$ according to $M(x) = 1$ iff $x \in X_m$, namely, flag $M$ determines if a state is marked in $\mathbf{SUP}$. Also, define $T: X \to \{1,0\}$ according to $T(x) = 1$ iff $(\exists s \in \Sigma^*)\xi(x_0, s) = x \,\&\, \delta(q_0, s) \in Q_m$, namely, flag $T$ determines if some corresponding state is marked in $\mathbf{G}$. Let $\mathcal{R} \subseteq X \times X$ be the binary relation such that for $x, x' \in X$, $(x, x') \in \mathcal{R}$. $x$ and $x'$ are called control consistent, if

$$E(x) \cap D(x') = E(x') \cap D(x) = \emptyset, \tag{5}$$

$$T(x) = T(x') \Rightarrow M(x) = M(x'). \tag{6}$$

While $\mathcal{R}$ is reflexive and symmetric, it need not be transitive, consequently it is not an equivalence relation.

Now, define $D^k: X \to Pwr(\Sigma_c^k)$ as $D^k(x) = \{\sigma \in \Sigma_c^k | \neg \xi(x, \sigma)! \,\&\, (\exists s \in \Sigma^*)[\xi(x_0, s) = x \,\&\, \delta(q_0, s\sigma)!]\}$. Let $\mathcal{R}^k \subseteq X \times X$ be the binary relation such that for $x, x' \in X$, $(x, x') \in \mathcal{R}^k$. $x$ and $x'$ are called control consistent w.r.t. $\Sigma_c^k$, if

$$E(x) \cap D^k(x') = E(x') \cap D^k(x) = \emptyset, \tag{7}$$

$$T(x) = T(x') \Rightarrow M(x) = M(x'). \tag{8}$$

Each pair of control consistent states $x, x' \in X$ in $\mathcal{R}$, is a member of $\mathcal{R}^k$, but the reverse is not true. We define a pair of exclusive control consistent states, in which a pair of states $x, x' \in X$ belongs to $\mathcal{R}^k$, but cannot belong to $\mathcal{R}$.

*Definition 1(Exclusive Control Consistent-ECC):* A pair of states $x, x' \in X$ are exclusive control consistent w.r.t. $\Sigma_c^k$, if

$(i)\ (x, x') \notin \mathcal{R}, (x, x') \in \mathcal{R}^k,$
$(ii)\ (\exists s, s' \in \Sigma^*)(\exists \sigma_i \in \Sigma), x = \xi(x_0, s), \xi(x, \sigma_i)!,$ (9)
$x' = \xi(x_0, s'), \neg \xi(x', \sigma_i)!, q_i = \delta(q_0, s\sigma_i), q_j = \delta(q_0, s'\sigma_i) \Rightarrow q_i \neq q_j$

Informally, a pair of states $x, x' \in X$ is ECC, if $(i)$ $(x, x')$ is a member in $\mathcal{R}^k$ but is not a member in $\mathcal{R}$, $(ii)$ each pair of corresponding states in $\mathbf{G}$ cannot reach to a same state in $\mathbf{G}$. This fact leads to the following definition to construct a cover on $X$, where at least a pair of ECC states belong to one subset of the cover.

*Definition 2:* A cover $\mathcal{C}^k = \{X_{i^k}^k \subseteq X | i^k \in I^k\}$ is a control cover on $X$ w.r.t. $\Sigma_c^k$, if

$(\exists i^k \in I^k), (\exists x, x' \in X_{i^k}^k), x, x'$ are ECC, (10)

$(\forall i^k \in I^k) X_{i^k}^k \neq \emptyset \wedge (\forall x, x' \in X_{i^k}^k)(x, x') \in \mathcal{R}^k,$ (11)

$(\forall i^k \in I^k)(\forall \sigma \in \Sigma)(\exists j^k \in I^k)\left[(\forall x \in X_{i^k}^k)\xi(x, \sigma)! \Rightarrow \xi(x, \sigma) \in X_{j^k}^k\right],$ (12)

Where $I^k$ is some index set in $\mathcal{C}^k$.

Thus, $\mathcal{C}^k$ is constructed if, by (10) At least a pair of states can be found such that they satisfy (9) in at least one subset of the cover, (11) Each pair of states that reside in each cell of control cover must be control consistent, and (12) For every event $\sigma \in \Sigma$, all states that can be reached from any state in $X_{ik}^k$ by a one-step transition $\sigma$ must be covered by the same cell $X_{jk}^k$. Note that we construct $\mathcal{C}^k$ such that it is control congruence, i.e. the state set $X$ is partitioned to several subset of states $X_{ik}^k$.

Given $\mathcal{C}^k$ on $X$, based only on the control information of $\Sigma_c^k$, an induced generator $\mathbf{LOC}^k = (Z^k, \Sigma, \zeta^k, z_0^k, Z_m^k)$ is obtained by the following construction,

$$(i) z_0^k \in Z^k \text{ such that } x_0 \in X_{i_0^k}^k,$$
$$(ii) Z_m^k = \{z^k \in Z^k | X_{ik}^k \cap X_m \neq \emptyset\},$$
$$(iii) \zeta^k : Z^k \times \Sigma \to Z^k \text{ with } \zeta^k(z_i^k, \sigma) = z_j^k, \text{if} \quad (13)$$
$$(\exists x \in X_{ik}^k) \xi(x, \sigma) \in X_{jk}^k \, \& \, (\forall x' \in X_{ik}^k)$$
$$\left[\xi(x', \sigma)! \Rightarrow \xi(x', \sigma) \in X_{jk}^k\right].$$

We can obtain a set of induced generators $\mathbf{LOC} = \{\mathbf{LOC}^k | k \in \mathcal{K}\}$. Let $L(\mathbf{LOC}) := \cap \{L(\mathbf{LOC}^k) | k \in \mathcal{K}\}$ and $L_m(\mathbf{LOC}) := \cap \{L_m(\mathbf{LOC}^k) | k \in \mathcal{K}\}$. $\mathbf{LOC}$ is a solution to the distributed supervisory control problem with event reduction.

In the rest of the paper, we prove that each local controller which is constructed by aforementioned procedure has reduced number of events comparing to the reduced supervisor. In the following proposition, it is proved that each local controller has at least one self-looped event at one state.

*Proposition 1:* Let $\mathbf{G}$ be the non-blocking plant, which consists of components $\mathbf{G}^k, k \in \mathcal{K}$, $\mathbf{SUP}$ be the supervisor of $\mathbf{G}$, and $\mathbf{LOC}^k$ be a local controller corresponding to $\mathbf{G}^k$. If $\exists x, x' \in X_{ik}^k$, such that $x, x'$ are ECC and $X_{ik}^k$ is corresponding to state $z_i^k$ in $\mathbf{LOC}^k$, then there exists $\sigma_j \in \Sigma_c^j, j \neq k$ which is self-looped at $z_i^k$.

*Proof:* Assume $\exists x, x' \in X_{ik}^k$, such that $x, x'$ are ECC. According to (9), there exists $\sigma_j \in E(x) \cap D(x')$ or $\sigma_j \in E(x') \cap D(x)$. Assume $\sigma_j \in E(x) \cap D(x')$. Since $\sigma_j \notin E(x) \cap D^k(x')$, $\sigma_j$ is disabled at $x'$ in $\mathbf{SUP}$, but it cannot be disabled at $z_i^k$ in $\mathbf{LOC}^k$. On the other hand, $\sigma_j$ is defined at corresponding states $q_i, q_j$ in $\mathbf{G}$, and according to (9), $q_i \neq q_j$. Hence, $\sigma_i$ also appears as a transition from $z_i^k$ to another state and appears as a self loop transition at $z_i^k$. It cannot be true, because $\mathbf{LOC}^k$ is deterministic automaton. Thus, $\sigma_i$ must either become a self loop transition at $z_i^k$ or become a transition from $z_i^k$ to another state $z_j^k$.

Assume $\sigma_i$ is a transition from $z_i^k$ to another state $z_j^k$, corresponding to one state in $\mathbf{SUP}$. It means that $s\sigma_i$ and $s'\sigma_i$ are both generated in $\mathbf{LOC}^k$ so that they reach to states $q_i, q_j$ in $\mathbf{G}$, respectively, and $q_i = q_j$. But, we know that $q_i \neq q_j$. Thus, $\sigma_i$ cannot be a transition from $z_i^k$ to another state. Therefore, $\sigma_i$ is a self loop transition at $z_i^k$.
Similarly, the proposition can be proved for $\sigma_i \in E(x') \cap D(x)$.

□

In Proposition 1, since $\sigma_i$ is disabled at some states of the supervisor, its observation affects the consistency of the supervisor behavior. It means that, if $\sigma_i$ is not observed, then the supervisor makes inconsistent decisions.

In order to prove the main claim, we show that $\sigma_i$ appears as a self loop transition at all states of $\mathbf{LOC}^k$. Since $\mathbf{LOC}^k$ is not authorized to disable $\sigma_i$, it is sufficient to prove that $\sigma_i$ appears as a self loop transition at states, where it is enabled. In Lemma 1, we prove the claim in a special case.

*Lemma 1:* Let $\mathbf{G}$ be the non-blocking plant, which consists of components $\mathbf{G}^k, k \in \mathcal{K}$, described by closed and marked languages $L(\mathbf{G}), L_m(\mathbf{G}) \subseteq \Sigma^*$ and $\mathbf{LOC}^k$ be a local controller corresponding to each $\mathbf{G}^k$. Let $\sigma_0 \in \Sigma_c^j, j \neq k$, and an arbitrary string belongs to $L(\mathbf{LOC}^k)$ be as shown in Fig. 1. If $[(\exists s \in \Sigma^*), \delta(q_0, s\sigma_0)!, \neg\delta(q_0, s\sigma_1)!, \exists i, z_i = \zeta_L(z_{L,0}, s) \Rightarrow z_i = \zeta_L(z_i, \sigma_0)]$, then $[\forall i, \zeta_L(z_i, \sigma_0)! \Rightarrow z_i = \zeta_L(z_i, \sigma_0)]$.

*Proof:* Assume the set of states and strings, are shown in Fig. 1. Define $P: \Sigma^* \to \Sigma_0^*$, $\Sigma_0 = \Sigma - \{\sigma_0\}$. We should prove that each pair of states, where $\sigma_0$ occurs in between, can be considered one state. We know that $P(s\sigma_0\sigma_1 s'\sigma_0) = P(s\sigma_1 s')$. Since a language, constructed by local controllers in the plant, is same as the language of the monolithic supervisor (see (3) and (4)), we can extend the observability property of the monolithic supervisor to each local controller. Thus, from Fig. 1, we can write,

$\forall \sigma \in \Sigma, s\sigma_0\sigma_1 s'\sigma_0\sigma \in L(\mathbf{LOC}^k) \cap L(\mathbf{G}), s\sigma_1 s' \in L(\mathbf{LOC}^k), s\sigma_1 s'\sigma \in L(\mathbf{G}) \Rightarrow s\sigma_1 s'\sigma \in L(\mathbf{LOC}^k).$ (14)

The string which occurs in both the local controller and the plant is shown in the first term of the antecedent in (14). Since $s\sigma_1 \notin L(\mathbf{G})$, then $s\sigma_1 s' \notin L(\mathbf{G})$. Thus, (14) is true. Similarly,

$s\sigma_0\sigma_1 s'\sigma_0 \in L_m(\mathbf{LOC}^k) \cap L_m(\mathbf{G}), s\sigma_1 s' \in L(\mathbf{LOC}^k) \cap L_m(\mathbf{G}) \Rightarrow s\sigma_1 s' \in L_m(\mathbf{LOC}^k).$ (15)

Since $s\sigma_1 s' \notin L(\mathbf{G})$, then $s\sigma_1 s' \notin L_m(\mathbf{G})$. Thus, (15) is true. Therefore, $z_n$ and $z_{n+1}$ can be considered one state, where $\sigma_0$ is a self-looped transition. However, the proposed supervisor localization procedure considers the plant cyclic (the task of the plant is assumed to be cyclic), even if it is not cyclic, we assume $\mathbf{G}$ is a cyclic plant. Thus, for $\forall i, \zeta_L(z_i, \sigma_0)!$ we use the above argument and conclude that $z_i = \zeta_L(z_i, \sigma_0)$. Since $\mathbf{LOC}^k$ is not authorized to disable $\sigma_0$, it can be considered a self loop transition at other states, where $\sigma_0$ is not defined at corresponding states in $\mathbf{G}$, i.e. $(\exists s \in \Sigma^*)[\delta(q_0, s) = q \& z_i = \zeta_L(z_{L,0}, s) \& \neg\delta(q, \sigma_0)!] \Rightarrow z_i = \zeta_L(z_i, \sigma_0)$. Therefore, $\sigma_0$ is self-looped at all states in $\mathbf{LOC}^k$.

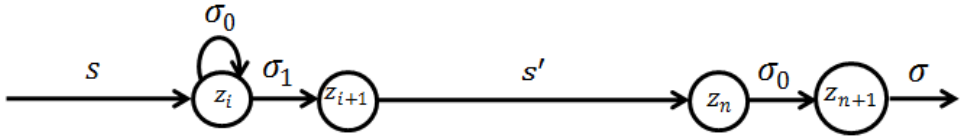

Fig. 1. A set of states and strings in $\mathbf{LOC}^k$, where $\sigma_0$ is self-looped at $z_i$

□

Now, we relax the assumption $s\sigma_1 \notin L(\mathbf{G})$, and show that Lemma 1 still does hold.

*Theorem 1:* Let $\mathbf{G}$ be the non-blocking plant, which consists of components $\mathbf{G}^k, k \in$

$\mathcal{K}$, described by closed and marked languages $L(\mathbf{G}), L_m(\mathbf{G}) \subseteq \Sigma^*$ and **SUP** be the supervisor of **G**. If $\exists x, x' \in X_{ik}^k$, such that $x, x'$ are ECC and $X_{ik}^k$ is corresponding to state $z_i^k$ in **LOC**$^k$, then $\exists \sigma_0 \in \Sigma_c^j, j \neq k$ such that $\sigma_0$ is self-looped at all states in **LOC**$^k$.

*Proof:* According to Proposition 1, $\sigma_0$ is self-looped at one state $z_i^k$. Following the proof of Lemma 1, if $s\sigma_1 \notin L(\mathbf{G})$, then (16) and (17) are satisfied, in Fig. 1, and $\sigma_0$ is self-looped at all states in **LOC**$^k$.

$s\sigma_0\sigma_1 s'\sigma_0\sigma \in L(\mathbf{LOC}^k) \cap L(\mathbf{G}), s\sigma_1 s' \in L(\mathbf{LOC}^k), s\sigma_1 s'\sigma \in L(\mathbf{G}) \Rightarrow s\sigma_1 s'\sigma \in L(\mathbf{LOC}^k),$ (16)

$s\sigma_0\sigma_1 s'\sigma_0 \in L_m(\mathbf{LOC}^k) \cap L_m(\mathbf{G}), s\sigma_1 s' \in L(\mathbf{LOC}^k) \cap L_m(\mathbf{G}) \Rightarrow s\sigma_1 s' \in L_m(\mathbf{LOC}^k).$ (17)

Now, we prove that (16) and (17) are true, even if $s\sigma_1 \in L(\mathbf{G})$. From Fig. 1, we know that $s\sigma_1 \in L(\mathbf{LOC}^k)$. There may be two cases related to $\sigma_1$: (a) $\sigma_1$ is a transition from $x$, or (b) $\sigma_1$ is a transition from $x'$, in **SUP**.

(a) Assume $\sigma_1$ is enabled at $x$. Then, either $\sigma_1$ is enabled at $x'$, or it is disabled at $x'$ in **SUP**. If $\sigma_1$ is disabled at $x'$ in **SUP**, then according to Proposition 1, it is self-looped at state $z_i^k$, where $\sigma_0$ is self-looped.

Now, assume that $\sigma_1$ is enabled at $x'$. Since **LOC**$^k$ is deterministic automaton, the next state from $x$ and $x'$ by transition $\sigma_1$ must be the same. Since $\sigma_1$ is a transition from one state in **LOC**$^k$, corresponding to a pair of control consistent states in **SUP**, to another state in **LOC**$^k$, corresponding to another pair of control consistent states in **SUP**, we can say **LOC**$^k$ can make decision without observing $\sigma_1$. It means that the next states by transition $\sigma_1$ must be merged with $x, x'$ (i.e. they belong to $X_{ik}^k$) and $\sigma_1$ is self-looped at state $z_i^k$, corresponding to subset of $X_{ik}^k$.

This argument can be continued for other subsequent states, until either all subsequent transitions become self-looped at the states belong to $X_{ik}^k$, (in this case, **LOC**$^k$ is reduced to one state with self-loop transitions, and the claim is proved), or some enabled events can be found in **LOC**$^k$, such that they are not defined at corresponding state in **G**. The latter was proved in Lemma 1.

(b) The proof is similar to case (a).

We summarize the above argument in the following statement,

If an event $\sigma_0$, which is not disabled in **LOC**$^k$, is self-looped at one state in **LOC**$^k$, then some other events may be self-looped there, until we can find an enabled event, such that it is not defined at corresponding state in **G**. According to Lemma 1, such events which are self-looped at one state in **LOC**$^k$, are also self-looped at all states of **LOC**$^k$. The proof is completed.

□

Similar to the method, proposed in [11], we prove that the local controllers, constructed by (10) - (13) are control equivalent to the monolithic supervisor w.r.t. the plant.

*Proposition 2:* Let **G** be the non-blocking plant, which consists of components $\mathbf{G}^k, k \in \mathcal{K}$, and **SUP** be the supervisor of **G**. If **LOC** is constructed by the procedure, proposed in (10) - (13), then **LOC** and **SUP** are control equivalent w.r.t. **G**, i.e. (3) and (4) are satisfied.

*Proof:* This proposition can be proved similar to Proposition 1 in [11].

□

However, we prefer a set of local controllers, with fewer states than the reduced supervisor; the localization procedure (presented in this paper)guarantees less number of events in each local controller comparing to the reduced supervisor. In the next section we illustrate the proposed method by an example.

## 4. Examples- Localization of supervisory control of transfer line with event reduction

Industrial transfer line consists of two machines $M_1$, $M_2$ and a test unit TU, that are linked by buffers $B_1$ and $B_2$ with capacities 3 and 1, respectively (Fig. 2). The DES model of industrial transfer line is shown in Fig. 3, where each subset of events are $\Sigma_{M_1} = \{1,2\}$, $\Sigma_{M_2} = \{3,4\}$, and $\Sigma_{TU} = \{5,6,8\}$, respectively. All events involved in the DES model are $\Sigma = \{1,2,3,4,5,6,8\}$, where controllable events are odd-numbered. If a work piece is accepted by TU, it is released from the system; if rejected, it is returned to $B_1$ for reprocessing by $M_2$. The specification is based on protecting $B_1$ and $B_2$ against underflow and overflow [19]. After synthesis the monolithic supervisor (Fig. 4), we obtain the reduced supervisor by **supreduce** procedure in TCT software [20] (Fig. 5). The original supervisor has 28 states and 65 transitions, and the reduced supervisor has 8 states and 31 transitions. Moreover, we construct local controllers for each component $M_1$, $M_2$ and TU, by the proposed method in this paper (Figs. 6-8). The tables of control data corresponding to the monolithic supervisor and each local controller are illustrated in tables 1-4. Control data are displayed as a list of supervisor states where disabling occurs, together with the events that must be disabled there.

In order to illustrate the extended method, it is sufficient to find a pair of states by which (9) is satisfied so that the state size of resulted local controller is less than the state cardinality of the reduced supervisor (Fig. 5). We check Proposition 1 for each local controller.

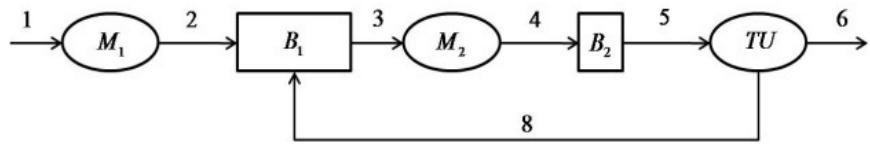

Fig. 2. Transfer Line

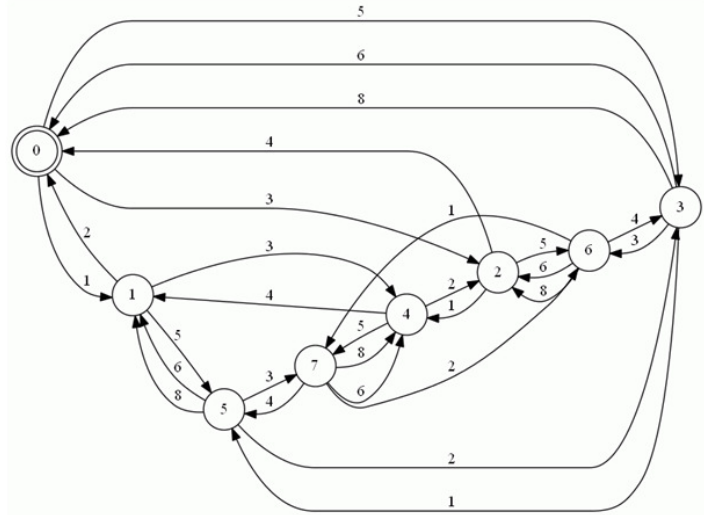

Fig. 3. TheDES model of Transfer Line

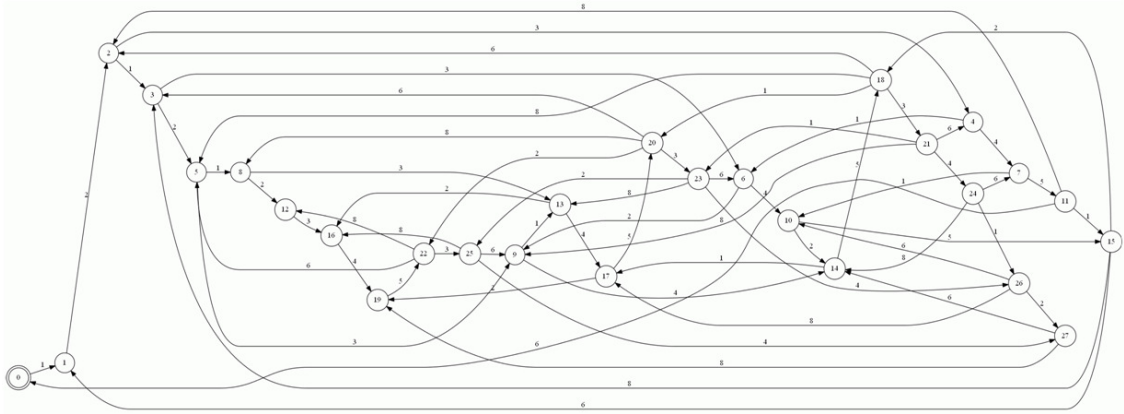

Fig. 4. Thesupervisor of Transfer Line

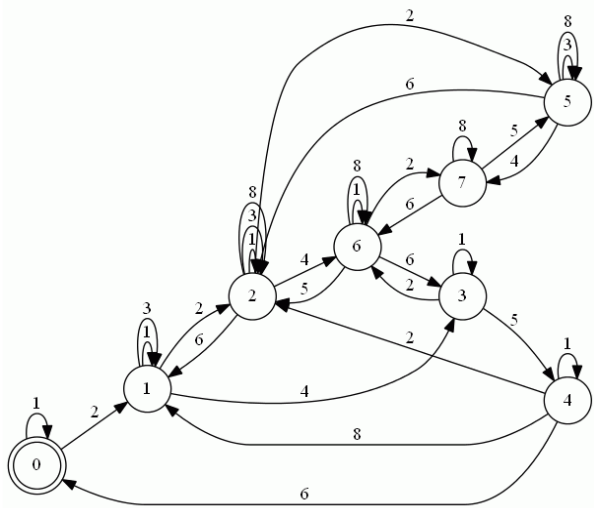

Fig. 5. Reduced supervisor of transfer line

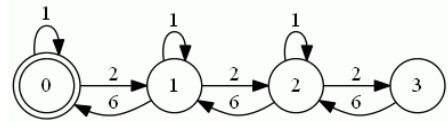

Fig. 6. Local controller $M_1$

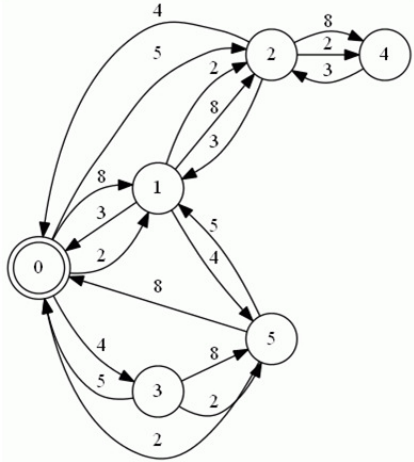
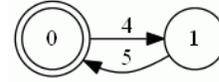

Fig. 7. Local controller $M_2$            Fig. 8. Local controller TU

Table 1. Control data of the monolithic supervisor

| State | Disabled Events | State | Disabled Events | State | Disabled Events |
|---|---|---|---|---|---|
| 0 | 3,5 | 8 | 5 | 16 | 1,5 |
| 1 | 3,5 | 9 | 5 | 17 | 3 |
| 2 | 5 | 10 | 3 | 19 | 1,3 |
| 3 | 5 | 11 | 3 | 22 | 1 |
| 4 | 5 | 12 | 1,5 | 24 | 3 |
| 5 | 5 | 13 | 5 | 25 | 1 |
| 6 | 5 | 14 | 3 | 26 | 3 |
| 7 | 3 | 15 | 3 | 27 | 1,3 |

Table 2. Control data for construction of the local controller $M_1$

| State | Disabled Events | State | Disabled Events | State | Disabled Events |
|---|---|---|---|---|---|
| 12 | 1 | 19 | 1 | 25 | 1 |
| 16 | 1 | 22 | 1 | 27 | 1 |

Table 3. Control data for construction of the local controller $M_2$

| State | Disabled Events | State | Disabled Events | State | Disabled Events |
|---|---|---|---|---|---|
| 0 | 3 | 11 | 3 | 19 | 3 |
| 1 | 3 | 14 | 3 | 24 | 3 |
| 7 | 3 | 15 | 3 | 26 | 3 |
| 10 | 3 | 17 | 3 | 27 | 3 |

Table 4. Control data for construction of the local controller TU

| State | Disabled Events | State | Disabled Events | State | Disabled Events |
|---|---|---|---|---|---|
| 0 | 5 | 4 | 5 | 9 | 5 |
| 1 | 5 | 5 | 5 | 12 | 5 |
| 2 | 5 | 6 | 5 | 13 | 5 |
| 3 | 5 | 8 | 5 | 16 | 5 |

Local controller $M_1$:

States 1 and 2 can be selected for checking Proposition 1 as follow,
$E(1) = \{2\}, D(1) = \{3,5\}$ and $E(2) = \{1,3\}, D(2) = \{5\} \Rightarrow 3 \in E(2) \cap D(1)$,

$(1,2) \notin \mathcal{R}$. Also, $D^1(1) = \{3,5\}$ and $D^1(2) = \{3,5\} \Rightarrow (1,2) \in \mathcal{R}^1$.

On the other hand, for strings $s\sigma_0 = 1, \mathbf{3}$, $s'\sigma_0 = 1,2,\mathbf{3}$, we have $x_1 = \xi(x_0, s), x_2 = \xi(x_0, s')$, $q_4 = \delta(q_0, s\sigma_0)$, $q_2 = \delta(q_0, s'\sigma_0)$, $q_4 \neq q_2$. Thus, event 3 is self-looped at all states of local controller $M_1$.

Local controller $M_2$:

States 11 and 19 can be selected for checking Proposition 1 as follow,
$E(11) = \{1,6,8\}, D(11) = \{3\}$ and $E(19) = \{5\}, D(19) = \{1,3\} \Rightarrow 1 \in E(11) \cap D(19)$, $(11,19) \notin \mathcal{R}$. Also, $D^2(11) = \{3\}$ and $D^2(19) = \{3\} \Rightarrow (11,19) \in \mathcal{R}^2$.

On the other hand, for strings $s\sigma_0 = 1,2,3,4,5, \mathbf{1}$, $s'\sigma_0 = 1,2,1,2,1,2,3,4, \mathbf{1}$, we have $x_{11} = \xi(x_0, s), x_{19} = \xi(x_0, s')$, $q_5 = \delta(q_0, s\sigma_0)$, $q_1 = \delta(q_0, s'\sigma_0)$, $q_5 \neq q_1$. Thus, event 1 is self-looped at all states of local controller $M_2$.

Local controller TU:

States 0 and 2 can be selected for checking Proposition 1 as follow,
$E(0) = \{1\}, D(0) = \{3,5\}$ and $E(3) = \{2,3\}, D(3) = \{5\} \Rightarrow 3 \in E(3) \cap D(0)$, $(0,3) \notin \mathcal{R}$. Also, $D^3(0) = \{5\}$ and $D^3(3) = \{5\} \Rightarrow (0,3) \in \mathcal{R}^3$.

On the other hand, for strings $s\sigma_0 = \mathbf{3}$ ($s = \epsilon$), $s'\sigma_0 = 1,2,1,\mathbf{3}$, we have $x_0 = \xi(x_0, s), x_3 = \xi(x_0, s')$, $q_2 = \delta(q_0, 3)$, $q_4 = \delta(q_0, 1,2,1,3)$, $q_2 \neq q_4$. Thus, event 3 is self-looped at all states of local controller TU.

Obviously, we see that each local controller has less number of states and less number of events, comparing to the reduced supervisor. Thus, we can say the supervisor of industrial transfer line is localizable in terms of both state reduction and event reduction criteria.

## 5. Conclusions

This paper addresses a new procedure for localization of the supervisory control, in which the local controllers have less number of events comparing to the reduced supervisor. In fact, this localization procedure not only provides fewer states, for easier implementation of each local controller on industrial systems, but also provides fewer events, in order to reduce the communication traffic between local controllers. Localizability is satisfied by state reduction in each local controller, whereas, event reduction guarantees reduction in communication traffic between local controllers. By event reduction property, we can investigate and compare the localizability and decomposability of a supervisor, in future work.

## References

[1] W. M. Wonham and P. J. Ramadge, "Modular supervisory control of discrete-event systems," Math. Control Signal Syst., vol.1, no.1, pp.13-30, 1988.
[2] G. Schafaschek, M. H. de Queiroz, J. E. R. Cury, "Local Modular Supervisory Control of Timed Discrete-Event Systems", WODES14, Cachan, May 2014, pp.271-277.
[3] L. Feng and W. M. Wonham, "Supervisory Control Architecture for Discrete-Event Systems", IEEE Trans. Autom. Control, vol.53, no.6, pp.1449-1461, 2008.
[4] Y. Willner and M. Heymann, "Supervisory control of concurrent discrete-event systems", International Journal of Control, vol.54, no.5, pp.1143-1169, 1991.
[5] F. Lin and W. M. Wonham, "Decentralized control and coordination of discrete-event systems with partial observation", IEEE Trans. Autom. Control, vol.35, no.12, pp.1330-1337, 1990.

[6] S. Mohajerani, R. Malik and M. Fabian,"Compositional synthesis of supervisors in the form of state machines and state maps", Automatica, vol.76, pp.277-281, 2017.
[7] R. Malik and M. Teixeira, "Modular Supervisor Synthesis for Extended Finite-State Machines Subject to Controllability", WODES16, Xi'an, June 2016, pp. 91-96.
[8] S. Mohajerani, R. Malik and M. Fabian, "A Framework for Compositional Synthesis of Modular Non-blocking Supervisors", IEEE Trans. Autom. Control, vol.59, no.1, pp.150-162, 2014.
[9] A. Vahidi, M. Fabian, B. Lennartson, "Efficient supervisory synthesis of large systems", Control Engineering Practice, vol.14, no.10, pp.1157-1167, 2006.
[10] R. Su and W. M. Wonham, "Supervisor reduction for discrete-event systems", Discrete-event Dyn. Syst., vol.14, no.1, pp.31–53, 2004.
[11] K. Cai and W. M. Wonham, "Supervisor Localization: A Top-Down Approach to Distributed Control of Discrete-Event Systems", IEEE Trans. Autom. Control, vol.55, no.3, pp.605-618, 2010.
[12] R. Zhang, K. Cai, W. M. Wonham, "Supervisor localization of discrete-event system s under partial observation", Automatica, vol.81, pp.142-147, 2017.
[13] K. Cai, R. Zhang, W.M. Wonham, "Relative Observability of Discrete-Event Systems and its Supremal Sublanguages", IEEE Trans. Autom. Control, vol.60, no.3, pp.659- 670, 2015.
[14] K. Cai and W. M. Wonham, "New results on supervisor localization, with case studies", Discrete-event Dyn. Syst., vol.25, no.1, pp.203-226, 2015.
[15] K. Rudie and W. M. Wonham, "Think globally, act locally: decentralized supervisory control", IEEE Trans. Autom. Control, vol.37, no.11, pp.1692-1708, 1992.
[16] F. Lin, W. M. Wonham, "On observability of discrete-event systems", Information Sciences, vol.44, no.3, pp.173-198, 1988.
[17] R. Cieslak, C. Desclaux, A. S. Fawaz, P. Varaiya, "Supervisory control of discrete-event processes with partial observations", IEEE Trans. Autom. Control, vol.33, no.3, pp.249-260, 1988.
[18] C.G. Cassandras and S. Lafortune, "Introduction to Discrete-event Systems", 2nd edn. Springer, NewYork, 2008.
[19] W. M. Wonham, "Supervisory control of discrete-event systems", 2016, http://www.control.utoronto.ca/DES.
[20] W. M. Wonham, Control Design Software: TCT. Developed by Systems Control Group, University of Toronto, Canada, 2014, http://www.control.utoronto.ca/cgi-bin/dlxptct.cgi.